# Entropy-based disciplinarity indicator: role taxonomy of journals in scientific communication systems and isolation degree. Knowledge importation/exportation profiles from journals and disciplines.

Jorge Mañana-Rodríguez*

**Abstract**

In this research, a new indicator of disciplinarity-multidisciplinarity is developed, discussed and applied. EBDI is based on the combination of the frequency distribution of subject categories of journals citing or cited by the analysis unit and the spread and diversity of the citations among subject categories measured with Shannon-Wiener entropy. Its reproducibility, robustness and consistence are discussed. Four of the combinations of its values when applied to the cited and citing dimensions lead to a suggested taxonomy of the role that the studied unit might have in terms of the transformation of knowledge from different disciplines in the scientific communication system and its position respect a hypothetical thematic core of the discipline in which it has been classified. The indicator is applied to the journals belonging to the first quartile of JCR-SSCI 2011 Library and Information Science and an indicator-based taxonomy is applied and discussed, pointing to differential thematic roles of the journals analyzed.

* Centre for Human and Social Sciences, CSIC.
jorge.mannana@cchs.csic.es /jorgemarod@gmail.com

## Introduction

In recent decades, the study of multidisciplinarity / disciplinarity has emerged as a core topic in science and technology studies and information & library science. IDR, Inter Disciplinary Research (Wagner et al. 2011), is a key aspect both for policymakers and researchers (National Academies, 2005). The increase in the number of journals has been observed, yet not exponentially (Mabe & Amin, 2001), having as one of the possible factors, as Ziman (1980) puts it *The impression of excessive proliferation arises mainly from the differentiation of journals to accommodate rapid expansion in specialized fields of research.*

A wide number of procedures and approaches have been designed in order to define and measure interdisciplinarity or disciplinarity as well as related concepts such as cross-disciplinarity (Porter & Chubin, 1985: COC and citation dispersal, continued by Tomov & Mutafov, 1996, developers of $I_{id}$ indicator) or trans-disciplinarity (Croyle, 2008, Hall et al. 2008). The methods used range from using pre-defined classifications (*classification-based*), mainly Thomson Reuters ISI Subject Categories (SC hereafter), i.e. van Raan & van Leeuwen (2002), to the development of procedures for *ad hoc* classification relying on factorization and clustering methods or *bottom-up approach* (i.e. Leydesdorff, 2007 or Rafols and Meyer 2010). The studies in this topic used approaches in terms of task (Porter and Rossini, 2007), process, and / or product and regarding the sources and nature of information, different inputs have been used, such as: word co-occurrence and neural networks, (Tijssen, 1992, or Zitt & Bassecoulard, 2006), co-classification of journals in pre-defined systems (Morillo, Bordons & Gómez, 2003) or citations (most of the studies rely on this latter aspect, such as Leydesdorff & Schank, 2008, Rafols & Meyer, 2010; Stirling, 2007 or Leydesdorff, & Rafols, 2011), since the interest in the cited work by the citing author is understood as underlying the citation process (Atkinson, 1984).

A recent work by Leydesdorff and Rafols (2011) compares the performance of various indicators aimed at the measurement of interdisciplinarity, among which there are network-based indicators such as Betwenness centrality ( Leydesdorff, 2007) and Rao-Stirling measures (Rao, 1982 and Stirling, 2007), and vector based indicators (Gini Index and Entropy), applied to a set of journals in JCR.

It might be possible to develop a disciplinarity-multidisciplinarity indicator which could be calculated only with the information associated to the studied unit (a vector based indicator by contrast with network-based indicators), susceptible of being applied to different nature units (i.e. journals, disciplines or individual authors). All these being factors which could involve a high practical "usability" by policymakers and/or other audiences, it remains unknown whether the combinations of the values of a vector-based indicator are interpretable in terms of the different combinations of values it can take when applied to the input (citing) and output (cited) dimensions, and if it could help creating a preliminary "taxonomy" of the role each unit plays in the disciplinary system in which it operates.

**Objectives:**

1 In this work the author seeks for a vector based indicator which values attempt to capture the degree of disciplinarity or multidisciplinarity (the latter corresponding to the definition in Rafols & Meyer, 2010) both in citing and cited dimensions using the minimum necessary information.

1.1 Since it is not the main aim of the indicator to assess either the quality or the robustness of the thematic classification system in which it is applied, it is a classification-based indicator and should be fully and easily reproducible in order to be of practical use by experts and non experts.

1.2 The indicator should be reliable, robust and consistent and its validity susceptible to be tested by multiple audiences.

2 The indicator should allow for taxonomy of journals according to the combination of its values in the citing and/or cited dimensions, and be helpful in placing a given journal close to or in the periphery of the thematic core of the subject category in which it is classified.

3 The indicator should be susceptible of being applied to different sets of journals regardless their thematic affiliation.

**Assumptions and development of the indicator:**

*Assumptions:* The percentage of citations coming or directed to journals classified in the same SC as the citing or cited journal (internal citations hereafter) is a proxy for a journal disciplinarity in the SC in which it has been classified: a journal citing or cited only by journals in its own SC is extremely monodisciplinary (Leydesdorff & Rafols, 2011, when referring to self-citations). It remains problematic to determine up to which point the percentage of citations coming or directed to journals not classified in the same SC as the studied journal (external citations hereafter) can be synthesized into a measure in which both, diversity and evenness or unevenness of the distribution are captured for a single journal. The external citations can come from a variable number of different SC, and the distribution of citations of the external citations among these variable numbers of SC may also be variably uneven.

*Shannon entropy as percentage of $H_{max}$*: A suitable indicator which captures both diversity or degree and evenness in the frequency distribution is Shannon's entropy. This has been recently discussed by Leydesdorff and Rafols (2011), who propose the use of the "local" entropy associated to the distribution of a given unit of analysis as a percentage of the maximum entropy of the distribution ($H_{max}$, log n) in order to avoid its size sensitivity when used as a raw measure. This approach has been used here, taking as n the maximum number of possible SCs, instead of the maximum number of citing journals, which was the basis for $H_{max}$ calculation used by the aforementioned authors; the use of frequency of publication in the same vs. other SC's as fundamental piece of information for individual researchers' interdisciplinary profile assessment has been widely developed in Porter et al 2007.

The entropy associated to a frequency distribution can be formulated as follows

$$H = -\sum_{i=1}^{n} p_i \ \ln\ (p_i)$$

Where $p_i = \frac{x_i}{X}$

And $X = \sum_{i=1}^{n} x_i$

H increases both with diversity of SC's in the external citations and with the evenness of the distribution. For a given journal, the frequency distribution of citations (*from* or *to* that unit) has an associated entropy, which interpretation in terms of information could be exemplified as follows: if a journal is cited only from journals belonging to one single SC in the classification system used, the uncertainty in that distribution is minimum as it is its associated entropy (Leydesdorff and Rafols (2011) Op. Cit.). Changes in both the number of SC's from which the citations come from, and unevenness in the frequency distribution would increase the uncertainty in the distribution, up to a maximum in which the uncertainty would be maximum as well as the entropy associated. This maximum uncertainty corresponds to $\ln n$ where $n$ is, in this case, the total number of possible SC from which the citations can come or be directed to.

Considering only the sources of the external citations, the entropy associated to the citation frequency distribution among SC's: [**statement 1**] increases when more SC's are added into the formula (a bigger "raw diversity"), and [**statement 2**] decreases with the unevenness of that distribution. Both properties reflect the degree of multidisciplinarity: this affirmation is maybe clearer for **[1]** than for **[2]**. It seems understandable that the more diversity in the SC sources of external citations involves greater "multi"-disciplinarity (supposing a comparative situation in which the percentage of internal citations is the same for both units) (Table 1).

In the case of **[2]**, a strongly uneven distribution involves a stronger citation pattern between the studied unit and certain specific disciplines rather than with others. Supposing two journals with equal percentage of internal citations, the same number of external citations and the same number of SC's from which those external citations come from or are directed to, the only fact which could differentiate them in terms of their degree of disciplinarity is the evenness or unevenness of their external citation pattern. If journal A has the same frequency of external citations along SC's, while B's frequency distribution of citations along SC's is strongly uneven, B could be considered more disciplinary, since it is more strongly related to *fewer* SC's than A.

**Table 1. Interpretation of changes in the frequency distribution of external citations and associated entropy values.**

| *Change in frequency distribution of external citation* | *Change in entropy values* | *Interpretation* |
|---|---|---|
| Δ Number of SC's involved (Δ in the diversity of sources) | ΔH | Δ Associated multidisciplinarity |
| Δ Unvenness of the distribution of citations among SC'S | - ΔH | - ΔH associated multidisciplinarity |

*Indicator formulation:* Following the reasoning explained, the following indicator is proposed:

Entropy-Based Disciplinarity Indicator (EBDI):

$$EBDI = \frac{\%IC}{\%H_{MAX(EC)+1}}$$

$\%IC$ is the percentage of citations from or to journals classified at least in the same subject category as the unit on which the indicator is being calculated (percentage of internal citations).

$\%H_{MAX(EC)+1}$ is the percentage that the entropy associated to the distribution of external citations represents respect the maximum entropy associated to the distribution of external citations (ln n, n being the maximum number of possible SC's in the system).

*Indicator interpretation:* If all citations are internal, the indicator value is equal to the percentage of internal citations, its value is 100, indicating extreme monodisciplinarity behavior of the studied unit in the discipline in which it has been classified. Co-classification of the journal in two disciplines might yield opposite results in each one. If all citations are external, then the indicator value is 0, indicating extreme multidisciplinarity (and maybe a problematic SC classification which might be revised by database managers).

The interpretation of the combination of high or low (the cutting point for high and los levels considered here is the percentile 50 in the rank-ordered distribution of EBDI values) levels of disciplinarity in the citing and cited dimensions is interpreted in this work as explained in table 2.

**Table 2: suggested taxonomy of studied units' role according to the combination of categorized levels of EBDI.**

| *Degree of disciplinarity:* ***citing*** *dimension.* | *Degree of disciplinarity:* ***cited*** *dimension.* | *Suggested interpretation of the "role" of the journal in the Sc* |
|---|---|---|
| HIGH | HIGH | The journal research front is clearly in its own SC and transforms knowledge from its own SC into knowledge which is mainly interesting for specialists in its own SC, adding a time differential. The journal might be at the disciplinary core of its SC. |
| LOW | HIGH | The journal takes knowledge from other SCs and transforms it into an output which is mainly interesting for researchers publishing in its discipline. It might be a form of knowledge input multidisciplinarity. |
| HIGH | LOW | The journal takes knowledge from its own SC and transforms it into an output which is mainly interesting for researchers publishing in other SCs journals. It might be a form of knowledge output multidisciplinarity. |
| LOW | LOW | The journal research front is not clearly in that SC. It takes knowledge from SCs other than the one in which it is classified and transforms it in an output which is also highly relevant to researchers publishing in journals belonging to other SCs. Its thematic relation to the SC is rather tangential. |

The interpretation exemplified in this case is only valid for journals; other units would entail different interpretations (for example, in the case of a subject category, the "degree of disciplinary isolation" could be the axis for the interpretation).

## Methodology.

In order to test the indicator, it was selected a set of journals conformed by those belonging to the first quartile in the impact factor rank of JCR social sciences edition 2011 in Information science & library science SC. The reason for choosing this subject category for this exploratory test is the better knowledge of the author as well as potential readers of this research of the nature of the publications of this field. The reason for choosing only the journals in the first quartile is the fact that the number of citations could be bigger than in other quartiles (since impact factor and raw number of citations have been found to be positively correlated in other studies, REF), and this could lead to meaningful conclusions in this first experimental application.

The abbreviated title of the journals citing and cited by the set of journals belonging to the first quartile in the IF ranking of Information science & Library science (2011) is available in Thomson Reuters JCR, and disaggregated by year of publication of the citing or cited item, as well as in the field "all years". This latter is the number of citations used here, though further studies could involve the temporal latency of citation disaggregating the different citations in each of the 10 years covered. These abbreviated titles where crossed with the full title and this with the subject category (or categories) in which citing and cited journals are classified using JCR master lists information. For this experimental application, only SC's of Social Sciences have been used.

Co-classification of journals is a key issue for defining internal or external citations has been treated as follows. Since a journal can be classified in various disciplines, in all cases in which the citing or cited journal is classified in the same SC as the unit of analysis it has been counted as an internal citation, while in the other cases it has been classified as an external citation. If the journal on which the indicator is being applied is classified in Information Science & Library Science and it gets 17 citations from journal B which is classified in Information Science & Library Science as well as in Geography, those citations will be counted as internal citations (Boolean "or"). The frequency of citations from or to journals belonging to different SC has been quantified. Then, over this information, in the cited and citing dimensions the percentage of internal citations, sum of external citations, associated entropy to the frequency distribution of external citations among SCs, the percentage that the latter "local" entropy represents respect the maximum entropy, the EBDI indicator and the "raw diversity", it is, the minimum number of different disciplines from which the citations come from have been also calculated (Table 3, example for the journal MIS Quarterly).

**Table 3: example of the data used for the calculation of EBDI for the journal MIS Quarterly.**

| *Value* | *CITED* | *CITING* |
|---|---|---|
| %INTERNAL CITATIONS | 52,58 | 37,04 |
| Σexternal citations | 1984 | 1438 |
| H | 2,03 | 1,99 |
| Hmax | 3,98 | 3,98 |
| %Hmax | 51,06 | 50,05 |
| EBDI | 1,01 | 0,73 |
| RAW DIVERSITY | 26 | 22 |

## Results

The results of the application of EBDI to the set of journals studied are plotted in chart 1. The citing and cited dimension thresholds which determine “high” or “low” levels of disciplinarity are the percentile 50 in the rank ordered EBDI distribution for each dimension, cited or citing.

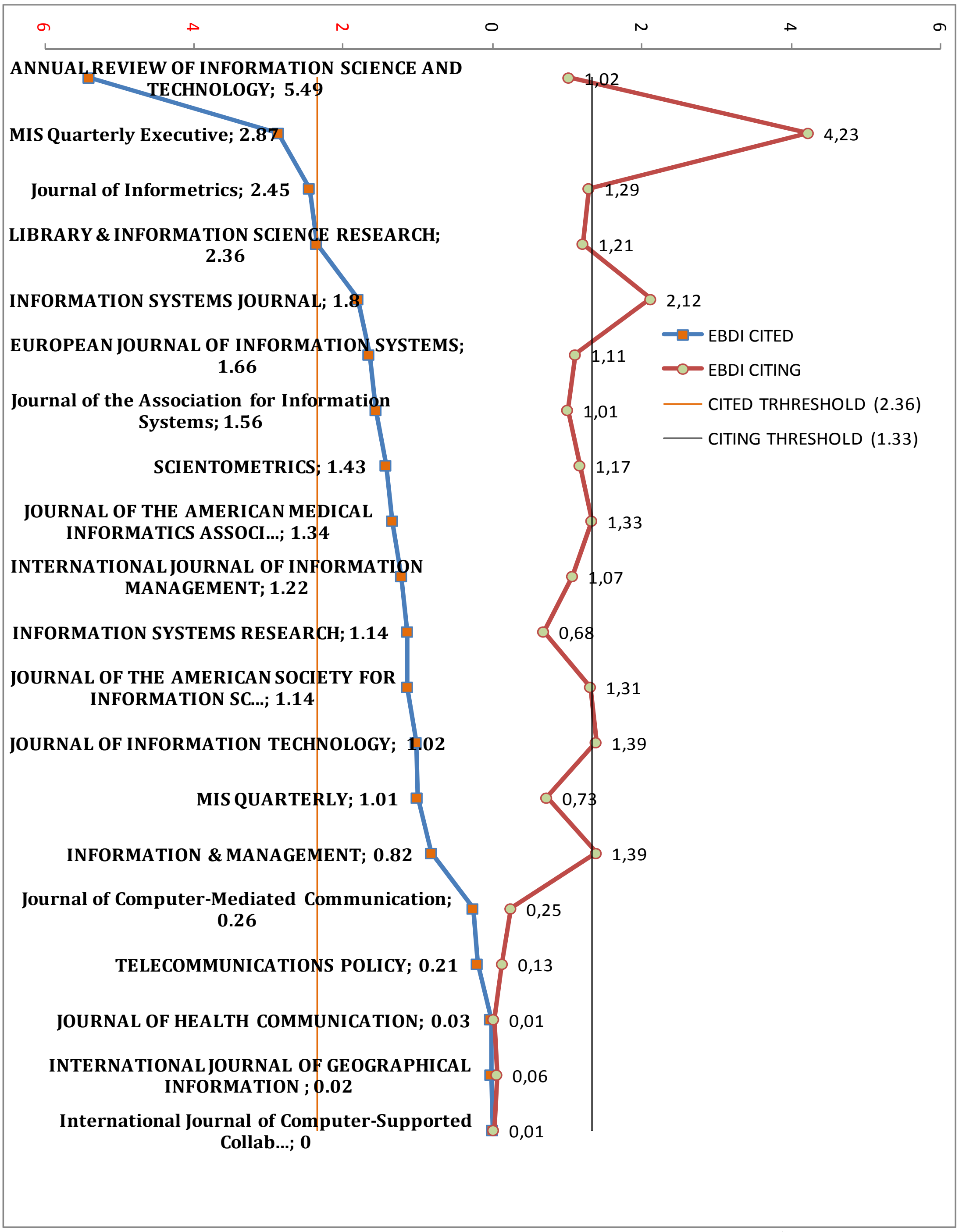

**Chart 1: Cited and citing indicators associated to journals in the 1st quartile (FI) of Information Science & Library Science SC in JCR Social Sciences Edition 2011**

As it can be observed in Chart 1, the Annual Review of Information Science and Technology is a clear case of knowledge input multidisciplinarity; papers whose journals are classified in a wide variety of SC's are cited by the authors publishing in this journal, but the papers published by the journal are mainly cited by authors publishing in other Information Science & Library Science journals: it might be a "knowledge importer".

MIS Quarterly Executive is highly disciplinary in both dimensions, cited and citing. Coherently with previous studies' results using other indicators (Leydesdorff and Rafols, 2011, Op. Cit.), this journal is highly monodisciplinary and might be taken as a core journal, from a knowledge classification perspective. It takes knowledge from its SC and transforms it into something mainly interesting for researchers publishing in that SC.

The Journal of Informetrics, though its values in the citing and cited dimensions are very close to the quartile threshold, could be accommodated in the taxonomy as a "knowledge importer", as in the case of the Annual Review of Information Science and Technology, since it is highly disciplinary in the cited dimension, but multidisciplinary in the citing dimension.

Information Systems Journal might be a clear example of "knowledge exporter", since it is disciplinary in the cited dimension but multidisciplinary in the citing dimension. The journal knowledge input frontier is mainly in its own SC, but the research published in the journal becomes interesting for researchers in other disciplines. In the same profile, it is possible to identify the Journal of Information Technology and the journal Information & Management.

Finally, among the journals which adscription to the field might be put into question, it is possible to mention the Journal of Health Communication and the International Journal of Geographical Information, which occupy rather remote positions to the disciplinary core journals' values of the indicator in terms of input and output disciplinarity.

Regarding the correlations between EBDI in the cited and citing dimensions, as well as other relevant correlations with indicators in the JCR register, the following Table 4 resumes some of the significant Spearman rank correlations of the indicator in the citing and cited dimensions at the .05 p-value level (n=20).

**Table 4: Significant Spearman rank correlations for EBDI in citing and cited dimensions, 5 year impact factor and article influence score.**

| *EBDI* | | *5Y IMPACT FACTOR* | *ARTICLE INFLUENCE SCORE* | *Citing EBDI* |
|---|---|---|---|---|
| Citing EBDI | ρ | -,457* | -,507* | 1 |
| | Sig. (2-tailed) | 0,05 | 0,03 | |
| Cited EBDI | ρ | | | ,492* |
| | Sig. (2-tailed) | | | 0,03 |

It can be observed a significant, moderate and negative correlation of EBDI in the citing dimension with the 5 year impact factor, which is not the case for the 3 year impact, or in the case of the cited dimension. There is also a significant, moderate and negative correlation of EBDI (citing) with the article influence score. Assuming that the indicator, when applied to the citing dimension, increases with the disciplinarity of the journal, this seems to be a negative factor regarding the long-term impact factor, as well as the article influence score. It is not surprising that the correlation between the two "sides" of the indicator is just moderate, since there are few journals with a balance between the number of different SCs from which the papers published in the journal gather the knowledge expressed as citations and the evenness in that distribution, and the same variables in the case of the citations the papers published in the journal receive. That lack of strong coherence between the two dimensions is a precondition for taxonomy.

**Discussion**

Any given indicator should be subject of analysis from a "desirable properties" perspective, both in general terms and when applied with bibliometric purposes. These properties are to be discussed in this section, together with potential weaknesses and applications.

*Indicator properties:* The reproducibility of an indicator by a wide range of subjects (specialists or not) is a desirable characteristic (Archambault et al, 2009). It guarantees the possibility of exercising the criticism and empirical contrast which is at the basis of any scientific endeavor. Certain bibliometric indicators which require the whole network of citations to be reproduced (such as SJR1 and 2, from Scimago Journal Rank) cannot be reproduced without the whole set of citing-cited vectors contained in the citation matrix derived from the source database or at least large pieces of it which is a serious drawback both for its reproducibility and the uses of the indicator. On the other side of the reproducibility spectrum, the efforts made by Mr. Loet Leydesdorff, making publicly accessible both data, detailed procedures, explanations for the calculus and analysis of his network-based indicators and results or the indicator SNIP developed by Henk Moed (Moed, 2010 and 2011) are, by the opposite, clear examples of reproducible results.

Due to the simplicity and straightforward calculation of the proposed indicator and the availability of the information required for its calculus, its values can be checked by an ample audience with access to WoS data.

The robustness of an indicator has several definitions and approaches (Archambault et al. 2009, Op. Cit.) but as a common core of their content, a low sensitivity to outliers is related to robustness and vice-versa. The use of percentages in the numerator and the denominator is the basis for the robustness of this indicator, since any extreme value would affect the indicator in a limited and proportional amount. Apart from this, since there are no factors in the indicator, the multiplicative risk of overweighting outliers is not involved in its calculus.

The concurrent validity, it is, the coincidence or co-variance of the conclusions derived from the indicator and other constructs on which there is consensus is yet to be determined, though some coincidences in the results with previous studies have been pointed out.

An indicator is consistent, specifically when it is a bibliometric indicator of total performance and according to Waltman et al (2011), *if adding the same publication to two different sets of publications never changes the way in which the indicator ranks the sets of publications relative to each other.* According to Rousseau and Leydesdorff (2011), a bibliometric indicator is consistent if the rank of two sets, not necessarily with the same number of elements, does not vary when items with 0 values are introduced.

Regarding Waltman et al.'s definition when applied to the indicator developed here, if EBDI takes the general form EBDI=a/b+1, being *a* and *b* positive natural numbers, then

**[1]** $\mathbf{a/(b+1) > (a+n)/(b+1) \leftrightarrow n<0.}$

Where n is a percentage of internal citations and cannot take negative values. If n′ is the result of adding a natural positive number and the value of the indicator resulting from this operation is denominated EBDI′, then:

**[2]** $\mathbf{\forall\, n>0: n<n' \; \exists!\, EBDI'>EBDI}$ **(QED)**

And EBDI′ can not occupy a position smaller than EBDI in a decreasing ranking.

In the case of Rousseau and Leydesdorff definition, the indicator takes into account only values greater than 0.

*Limitations and shortcomings*

The implementation of the indicator in bigger and more diverse sets of journals might reveal weaknesses when comparing strongly different SC's and might require specific normalization procedures. Also, in this case, the indicator has not been applied to the citations in journals belonging to Science Citation Index, and it remains unknown how it would be possible to characterize a journal according to the citations in both dimensions and in the two branches of knowledge. The use of this indicator at the individual level or when applied to other kinds of units such as whole fields, might be of use, though it remains to be know if it would be necessary to modify the indicator in order to satisfy the characteristics of the studied unit or other specific targets. Finally, though some evidence of concurrent validity of the indicator has been explained in this article, it remains yet to be done a wide validation study.

*Possible applications and further research*

The classification of journals in subject categories remains a persistent problem all along the different systems. Citations from or to a journal have been used for decades as a basis, jointly with other variables, as reliable evidence towards a satisfying classification (whatever is the precise meaning of "satisfying" strongly depends on the needs the system tries to fulfill). The taxonomy here proposed, particularly in the case of low values of disciplinarity in cited and citing dimensions might be helpful as a possible evidence of weak thematic relationship to the SC in which the journal has been classified, thus possibly contributing to a more depurated classification system/procedure. In the opposite side of the spectrum, journals with high levels of disciplinarity in the two dimensions might be considered as a "reference point" from which distance measures could be taken. The implementation of entropy-based disciplinarity indicators in networks might also be a possible approach when the whole matrix of co-citations is to be characterized or studied. Among the possible further research possibilities of the indicator, it is possible to incorporate (since data is available) a time differential, it is to say, the diachronic study of disciplinarity tendencies along the ten year frame considered in JCR. The indicator, due to its immediate availability, might be of use for policymakers when taking decisions related to targets involving multidisciplinarity as a goal. Finally, the taxonomy of cited-citing disciplinarity and the associated roles might be an interesting segmentation variable for identifying specific needs of geographically, thematically or institutionally defined "journal populations" and their consumers, as well as for the commercial fulfillment of those needs such as, i.e. “knowledge importers” journals consumers or authors might find more attractive more variety in the thematic diversity of the journal packages commercially available than those publishing in “knowledge exporter” journals, while the latter might find more useful to count with a “thematic core” set of subscripted journals .

## Conclusions

The use of easily available data in a straight-forward computation indicator such as EBDI might be considered a desirable property. When applied to journals in the first quartile by impact factor of the Information Science and Library Science SC in JC, the Annual Review of Information Science and Technology has been positioned as the most disciplinary in the cited dimension, while MIS Quarterly is the most disciplinary in the cited dimension. On the opposite side of the spectrum of disciplinarity, the International Journal of Computer Supported Collaborative Learning is classified as the less disciplinary (or more multidisciplinary in the frame of the SC in which it has been classified) in both dimensions, cited and citing. The application of the indicator to a well known set of journals has resulted in a taxonomy which allows the categorization of journals according to their profile regarding their behavior in the cited and the citing dimensions. The results of the application of the indicator are fully reproducible and might be useful for the characterization of sets of journals according to the various profiles they can fit in, as well as for the study of the scientific production of units of study others than scientific journals, as well as for the segmentation of the population conformed by database and subscription consumers according to their potentially differential interests.

**ADDENDA IMPORTING/EXPORTING DISCIPLINES IN SOCIAL SCIENCES ACCORDING TO THEIR SPECIALIZATION PROFILE BY EBDI INDICATOR. KNOWLEDGE IMPORTATION/EXPORTATION BETWEEN SOCIAL SCIENCES AND SCIENTIFIC AND TECHNOLOGICAL FIELDS.**

As an addenda of a previous paper uploaded by Jorge Mañana Rodríguez regarding the development of the Entropy Based Disciplinarity Indicator and its application to journals in Information JCR Science & Library Science subject category, as well as the taxonomy of journals according to their knowledge importation/exportation profile (TU. 26/2013/02), in this section, the author provides graphs, tables and cases of calculation of EBDI (or other network indicators) applied, in this case at the discipline level to characterize Social Sciences as disciplinary (specialized or isolated or multidisciplinarity and its derived taxonomy) in the framework of their knowledge importation-exportation behavior both in the citing and/or cited dimensions. It also provides the information required to analyze their degree of knowledge exchange (importation/exportation of knowledge) with fields such as the sciences, technological fields of knowledge or the arts and the humanities.

1st: As a reminder of the original formulation of the indicator, the following explanation, extracted from the same authors' paper available at http://arxiv.org/abs/1302.6528, TU. 26/2013/02 is here shown.

Entropy-Based Disciplinarity Indicator (EBDI):

$$EBDI = \frac{\%IC}{\%H_{MAX(EC)+1}}$$

$\%IC$ is the percentage of citations from or to journals classified at least in the same subject category as the unit on which the indicator is being calculated (percentage of internal citations).
$\%H_{MAX(EC)+1}$ is the percentage that the entropy associated to the distribution of external citations represents respect the maximum entropy associated to the distribution of external citations (ln n, n being the maximum number of possible SC's in the system).
*Indicator interpretation:* If all citations are internal, the indicator value is equal to the percentage of internal citations, its value is 100, indicating extreme monodisciplinarity behavior of the studied unit in the discipline in which it has been classified. Co-classification of the journal in two disciplines might yield opposite results in each one. If all citations are external, then the indicator value is 0, indicating extreme multidisciplinarity (and maybe a problematic SC classification which might be revised by database managers).

In the following table, it is shown a taxonomy of psychology categories and other two control subject categories (Ethnic studies and Cultural studies).

**Table 1: Taxonomy of disciplines according to their Entropy-Based Disciplinarity Indicator (disciplinarity standing also for specialization) and the derived taxonomy of knowledge importation/exportation.**

| CITED EBDI VALUES | | CITING EBDI VALUES | | DIFFERENCE CITED-CITING | DISCIPLINE TYPE |
|---|---|---|---|---|---|
| PSYCHOLOGY MATHEMATICAL | 1,326 | PSYCHOLOGY MATHEMATICAL | 0,643 | 0,68 | knowledge importer |
| PSYCHOANALYSIS | 2,391 | PSYCHOANALYSIS | 1,823 | 0,57 | knowledge importer |
| PSYCHOLOGY DEVELOPMENTAL | 0,679 | PSYCHOLOGY DEVELOPMENTAL | 0,640 | 0,04 | knowledge importer |
| ETHNIC STUDIES | 0,31 | ETHNIC STUDIES | 0,34 | -0,03 | knowledge exporter |
| PSYCHOLOGY MULTIDISCIPLINARY | 0,304 | PSYCHOLOGY MULTIDISCIPLINARY | 0,370 | -0,07 | knowledge exporter |
| PSYCHOLOGY APPLIED | 0,440 | PSYCHOLOGY APPLIED | 0,517 | -0,08 | knowledge exporter |
| PSYCHOLOGY CLINICAL | 0,628 | PSYCHOLOGY CLINICAL | 0,719 | -0,09 | knowledge exporter |
| PSYCHOLOGY BIOGICAL | 0,636 | PSYCHOLOGY BIOGICAL | 0,749 | -0,11 | knowledge exporter |
| PSYCHOLOGY SOCIAL | 0,844 | PSYCHOLOGY SOCIAL | 1,051 | -0,21 | knowledge exporter |
| PSYCHOLOGY EXPERIMENTAL | 0,937 | PSYCHOLOGY EXPERIMENTAL | 1,159 | -0,22 | knowledge exporter |
| PSYCHOLOGY EDUCATIONAL | 0,740 | PSYCHOLOGY EDUCATIONAL | 1,273 | -0,53 | knowledge exporter |
| CULTURAL STUDIES | 0,83 | CULTURAL STUDIES | 1,47 | -0,64 | knowledge exporter |

The next graph shows the citation relationships established along disciplines of psychology with other fields of knowledge, according to JCR 2010 data (only the first 10 disciplines by rank-order of number of citations has been inputted in the network). The first network expresses the relationships in the cited dimension with social sciences fields, while the second involves only the relationships of psychology subject categories with fieds in the Science Citation Index, thus revealing its citation relationships with the scientific (no social sciences) and technological fields.

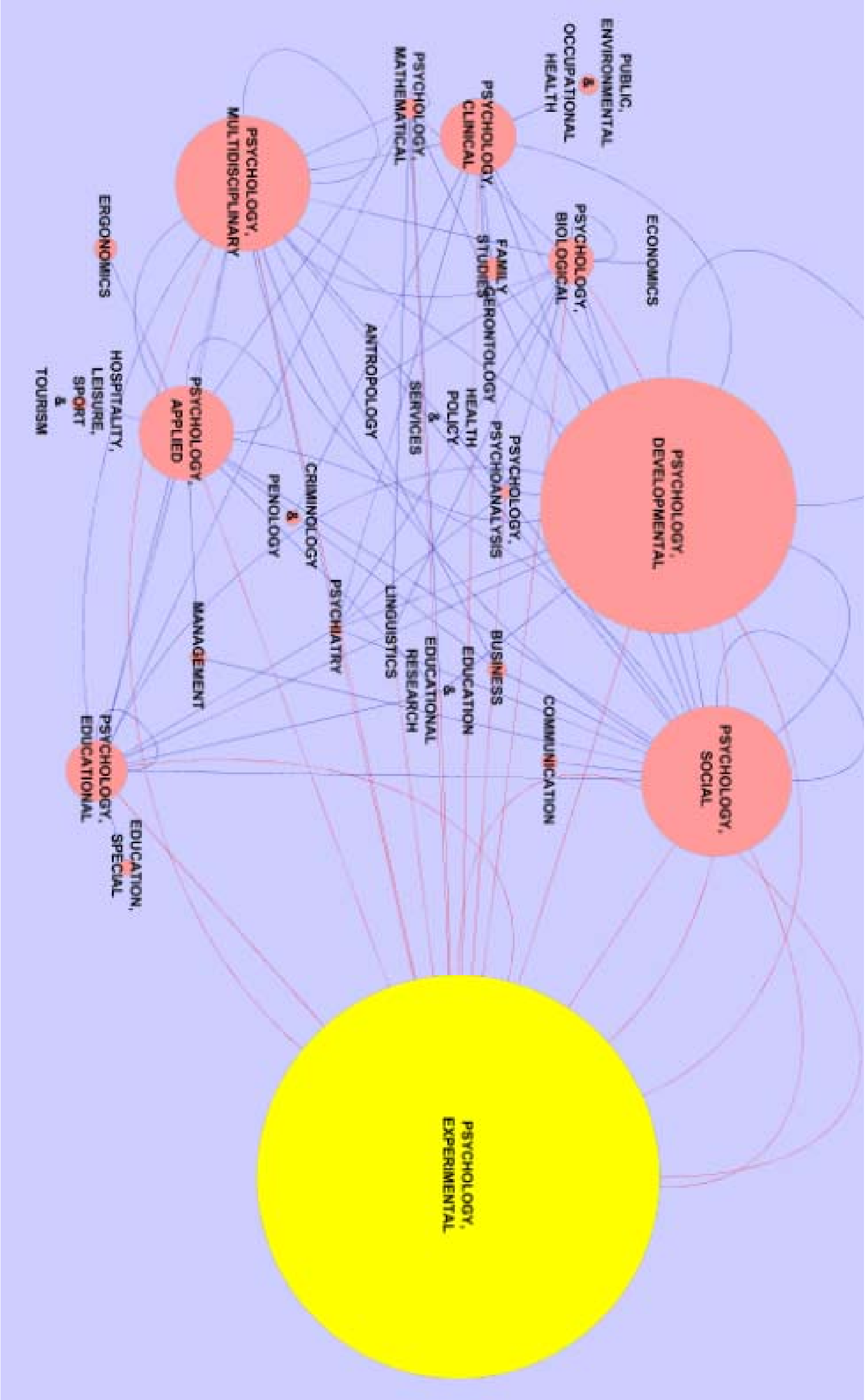

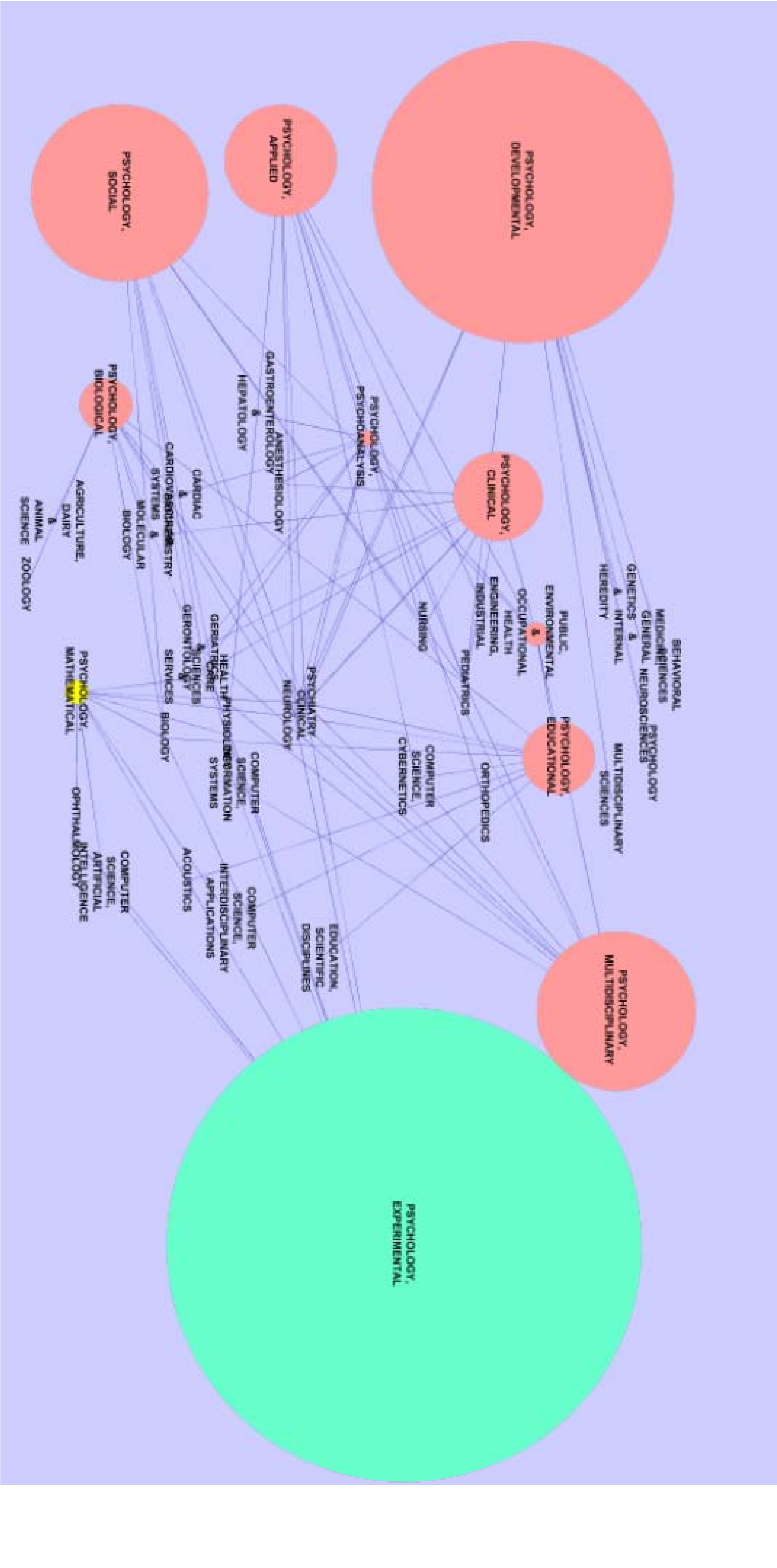
PSYCHOLOGY, SOCIAL
PSYCHOLOGY, APPLIED
PSYCHOLOGY, DEVELOPMENTAL
PSYCHOLOGY, BIOLOGICAL
GASTROENTEROLOGY & HEPATOLOGY
PSYCHOLOGY, PSYCHOANALYSIS
ANESTHESIOLOGY
PSYCHOLOGY, CLINICAL
CARDIAC & CARDIOVASCULAR SYSTEMS
BIOCHEMISTRY & MOLECULAR BIOLOGY
AGRICULTURE, DAIRY & ANIMAL SCIENCE
ZOOLOGY
GENETICS & HEREDITY
MEDICINE, GENERAL & INTERNAL
BEHAVIORAL SCIENCES
NEUROSCIENCES
PSYCHOLOGY
PUBLIC, ENVIRONMENTAL & OCCUPATIONAL HEALTH
ENGINEERING, INDUSTRIAL
NURSING
PEDIATRICS
GERIATRICS & GERONTOLOGY
HEALTH CARE SCIENCES & SERVICES
PHYSIOLOGY
BIOLOGY
PSYCHIATRY
CLINICAL NEUROLOGY
PSYCHOLOGY, MATHEMATICAL
PSYCHOLOGY, EDUCATIONAL
MULTIDISCIPLINARY SCIENCES
COMPUTER SCIENCE, INFORMATION SYSTEMS
COMPUTER SCIENCE, CYBERNETICS
ORTHOPEDICS
OPHTHALMOLOGY
COMPUTER SCIENCE, ARTIFICIAL INTELLIGENCE
ACOUSTICS
COMPUTER SCIENCE, INTERDISCIPLINARY APPLICATIONS
EDUCATION, SCIENTIFIC DISCIPLINES
PSYCHOLOGY, MULTIDISCIPLINARY
PSYCHOLOGY, EXPERIMENTAL

Finally, in the next table, it is shown the entropy associated to the external citations received by each discipline in the cited and citing dimensions, as the basis for the analisis of its specialization.

| **DIMENSION** | **DISCIPLINE** | **EXTERNAL CITATIONS' DISTRIBUTION ENTROPY** |
|---|---|---|
| CITED | CULTURAL STUDIES | 0,287 |
| CITED | ETHNIC STUDIES | 0,306 |
| CITED | PSYCHOLOGY, EXPERIMENTAL | 0,212 |
| CITED | PSYCHOLOGY, MATHEMATICAL | 0,268 |
| CITED | PSYCHOLOGY, BIOLOGICAL | 0,208 |
| CITED | PSYCHOLOGY, CLINICAL | 0,258 |
| CITED | PSYCHOLOGY, PSYCHOANALYSIS | 0,154 |
| CITED | PSYCHOLOGY, DEVELOPMENTAL | 0,256 |
| CITED | PSYCHOLOGY, SOCIAL | 0,284 |
| CITED | PSYCHOLOGY, EDUCATIONAL | 0,250 |
| CITED | PSYCHOLOGY, MULTIDISCIPLINARY | 0,303 |
| CITED | PSYCHOLOGY, APPLIED | 0,272 |
| CITING | CULTURAL STUDIES | 0,258 |
| CITING | ETHNIC STUDIES | 0,314 |
| CITING | PSYCHOLOGY, EXPERIMENTAL | 0,204 |
| CITING | PSYCHOLOGY, MATHEMATICAL | 0,224 |
| CITING | PSYCHOLOGY, BIOLOGICAL | 0,206 |
| CITING | PSYCHOLOGY, CLINICAL | 0,227 |
| CITING | PSYCHOLOGY, PSYCHOANALYSIS | 0,149 |
| CITING | PSYCHOLOGY, DEVELOPMENTAL | 0,235 |
| CITING | PSYCHOLOGY, SOCIAL | 0,256 |
| CITING | PSYCHOLOGY, EDUCATIONAL | 0,241 |
| CITING | PSYCHOLOGY, MULTIDISCIPLINARY | 0,290 |
| CITING | PSYCHOLOGY, APPLIED | 0,255 |